\documentclass[10pt,twocolumn]{article}

\usepackage[a4paper,left=53.798bp,right=39.334bp,top=53.798bp,bottom=124.58bp,columnsep=23.884bp]{geometry}
\usepackage{helvet}
\usepackage{graphicx}
\usepackage{amsmath,amssymb}
\usepackage{booktabs}
\usepackage{array}
\usepackage{enumitem}
\usepackage{caption}
\usepackage{titlesec}
\usepackage{indentfirst}
\usepackage{url}
\usepackage{balance}
\usepackage{algorithmic}
\usepackage{hyperref}

\hypersetup{hidelinks}
\setlist[itemize]{leftmargin=1.5em,itemsep=1pt,topsep=2pt}
\setlist[enumerate]{leftmargin=2.25em,itemsep=1pt,topsep=7.65pt}

\titleformat{\section}{\normalfont\fontfamily{ptm}\bfseries\fontsize{12}{14}\selectfont}{\thesection.}{1em}{\MakeUppercase}
\titleformat{\subsection}{\normalfont\fontfamily{ptm}\bfseries\fontsize{12}{14}\selectfont}{\thesubsection}{1em}{}
\titlespacing*{\section}{0pt}{8pt plus 2pt minus 1pt}{4pt}
\titlespacing*{\subsection}{0pt}{7pt plus 2pt minus 1pt}{3pt}

\renewenvironment{abstract}{%
  \noindent\fontfamily{ptm}\bfseries\fontsize{12}{14}\selectfont ABSTRACT\par\vspace{3.99pt}%
  \normalfont\fontfamily{ptm}\fontsize{10}{12}\selectfont\noindent\ignorespaces
}{\par}

\newcommand{\AAsys}{\mbox{A$^{A}$}}
\newcommand{\DT}{D_{T}}
\newcommand{\CR}{C_{R}}
\newcommand{\vect}[1]{\boldsymbol{#1}}
\newcounter{algocounter}

\makeatletter
\renewcommand{\@maketitle}{%
  \newpage
  \null
  \vskip 11.53pt%
  \begin{center}%
    {\sffamily\bfseries\fontsize{18}{21.6}\selectfont \@title \par}%
    \vskip 29.81pt%
    {\fontencoding{T1}\fontfamily{phv}\fontsize{12}{12.2}\selectfont \@author \par}%
  \end{center}%
  \vskip 39.52pt%
}
\makeatother

\title{Smartphone Audio Based Distress Detection}
\author{%
Anil Sharma, Sarthak Ahuja, Mayank Gautam, and Sanjit Kaul\\
IIIT-Delhi, India\\
\{anils, sarthak12088, mayank10047, skkaul\}@iiitd.ac.in}
\date{}

\begin{document}
\maketitle

\begin{abstract}
We investigate an unobtrusive and $24\times7$ human \emph{distress detection and signaling system, Always Alert}, that requires the smartphone, and not its human owner, to be on alert. The system leverages the microphone sensor, at least one of which is available on every phone, and assumes the availability of a data network. We propose a novel two-stage supervised learning framework, using support vector machines (SVMs), that executes on a user's smartphone and monitors natural vocal expressions of fear---screaming and crying in our study---when a human being is in harm's way. The challenge is to achieve a high distress detection rate while ensuring that the false alarm rate is a manageable overhead, while a typical smartphone user goes about living life as usual. We train the learning framework with carefully selected audio fingerprints of distress and of varied environmental contexts. The audio is used to tune the learning framework to obtain a \emph{desirable} distress detection rate and false alarm rate (FAR). The ability of the proposed framework to detect distress in rather challenging audio environments is demonstrated. Exploiting the time contiguous nature of false alarms further allows us to reduce the FAR. We show the feasibility of using our framework anytime and anywhere by testing it over many hours of audio fingerprints recorded by volunteers on their smartphones, as they went about their daily routines. We are able to achieve high distress detection rates at an average overhead that is equivalent to about 1 facebook post every 3 to 4 hours.
\end{abstract}

\section{Introduction}
Security systems (burglar alarms, intrusion detection systems) abound for residential and commercial establishments. A breach of security leads to an alarm that is followed by a visit from law enforcement agencies. Often, the visit is preceded by an attempt to confirm that a breach had indeed taken place, that is the alarm was not a false alarm~\cite{ref10}.

While crimes against individuals are probably as old as humanity, anytime anywhere security for a typical individual going about life as usual is fairly new. It has been made possible by smart devices (for example, smartphones and watches) and wide network coverage. However, the current approaches leave a lot to be desired. Most approaches require the smartphone (or device) owner to be alert enough to raise panic using the phone, often by what is equivalent to pressing a panic button. The human needs to be constantly on guard. While this is desirable when passing through a place known for its notoriety, it is unnatural and onerous to be on guard always, especially when in familiar places and amongst familiar faces~\cite{ref1,ref5}.

We propose Always Alert (\AAsys), a distress detection system, in which a person's smartphone is always on guard. \AAsys\ receives audio inputs from the phone's microphone and processes them to detect the presence of screaming and crying, which are known natural responses to distressful circumstances~\cite{ref13}. \AAsys\ needs to address two challenges before it can be deployed in the real world. Firstly, it must do its primary job of detecting distress with high accuracy. It must achieve a good accuracy in real world settings in which the background often contains other sounds. These other sounds form the environmental context of the phone (and its user). As a result, audio inputs received by a user's phone are not as clean as recordings made under controlled and quiet settings. Also, a user may place the phone at various locations, for example, in hand, or in a pocket, or in a bag, further affecting the quality of audio received by the phone's microphone. Secondly, it must trigger a very small number of false alarms. Specifically, the overheads on law enforcement due to false alarms must be minimal.

There are many other works on scream detection~\cite{ref18,ref31,ref35}. To our knowledge none of them address detection $24\times7$ over varying environmental contexts. Also, they often use fixed sensors and training over carefully chosen non-scream audio samples. For example, in~\cite{ref18} the authors use microphone arrays and train their learning framework to distinguish between a scream, and sounds created on applause, laughter, crying, glass break, and clap. In contrast, to exemplify the challenges that \AAsys\ must address, it must be able to detect screams when a microwave or any other home appliance is on in the background. Also, \AAsys\ must not falsely categorize sounds from a home appliance as screams and forward the false detections (alarms) to law enforcement.

\AAsys\ follows a four stage approach to distress detection. The first stage, \emph{Speech Filter}, detects distress with high probability. Speech Filter is trained using clean scream, crying, and normal speech samples. However, Speech Filter leads to a significant (about 10\%) false alarm rate. Audio that is accepted as a scream by Speech Filter is then processed by \emph{Context Filter}. This stage is trained to understand environmental contexts. It brings down the false alarm rate by an order of magnitude (to about 1\%) and has a negligible impact on the true scream detection rate. A FAR of 1\% is about 15 minutes of false alarms generated by a user of \AAsys\ over 24 hours. This is further reduced by \emph{Temporal Analysis}, which exploits the fact that false alarms often occur contiguously in time. The environmental context is responsible for them and often does not change for extended periods of time. All audio samples that pass the above stages as a distress sample are sent to friends-in-the-detection-loop before escalation to law enforcement. Evaluation of \AAsys\ using volunteer data suggests that the false alarms amount to an average overhead of one (facebook\textsuperscript{\copyright}) post every 3 to 4 hours to friends-in-the-detection-loop.

Our specific contributions are as follows.
\begin{enumerate}
\item To our knowledge, we are the first to propose an entirely smartphone audio based $24\times7$ distress detection system. All stages starting with recording of audio samples, classification, and raising of alarm, execute on the phone in real time. We provide an outline of the implementation of the application and a thorough evaluation of its suitability to execute on a smartphone.
\item We propose and evaluate a novel multistage distress detection and false alarm rejection framework that is able to achieve high distress detection rates and low false alarm rates, even in the presence of a variety of sounds from environmental contexts. The framework allows selection of a desirable tradeoff between false alarm rate (FAR) and detection rate, across a range of harsh environmental contexts.
\item Evaluation of the feasibility of the use of friends-in-the-detection-loop to further reduce the FAR so as not to create an unnecessary burden on law enforcement.
\item Extensive evaluation of our proposals on many hours of volunteer data that was collected by 16 volunteers using smartphones and going about their usual daily routine. For volunteer collected data we show that overheads due to false alarm rate are on an average equivalent to a facebook post every 3--4 hours.
\end{enumerate}

\begin{figure}[t]
\vspace{-8.80pt}
\centering
\includegraphics[width=198bp]{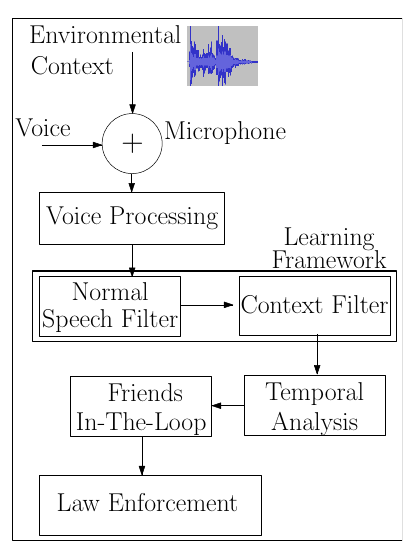}
\captionsetup{skip=9.2pt}
\caption{Architecture of \AAsys.}
\label{fig:architecture}
\vspace{21pt}
\end{figure}

Rest of the paper is organized as follows. Section~2 describes the architecture of \AAsys. Sections~3 and~4 describe data collection and processing. Section~5 describes our learning framework, which is followed by a description of temporal analysis in Section~6. Evaluation methodology is described in Section~7. Section~8 and~9 detail the performance evaluation of \AAsys. Section~10 describes our implementation on Android and its energy footprint. We discuss our limitations and a few possible approaches in Section~11. Related works are described in 12. We conclude in 13.

\section{Architecture}
Figure~\ref{fig:architecture} shows the various blocks that constitute \AAsys. Note that all of the blocks other than friends-in-the-loop and law enforcement execute as a smartphone application. \AAsys\ senses the environmental context and any human audio signals using a phone's one or more microphones. The resulting audio is sent through the voice processing block, which converts the input audio stream into smaller chunks, samples of 2 seconds length, that are each converted into (MFCC) feature vectors. The feature vectors are input to the learning framework that consists of the blocks Normal Speech Filter and Context Filter. The phone categorizes a sample to belong to the distress category only if both the Normal Speech Filter and the Context Filter categorize the sample as distress.

Once a sample is categorized as distress, the phone sets its state to \emph{alarm raised}. The alarm is sent to friend(s)-in-loop together with a snapshot of audio around the time of alarm, who for this study are subscribers to a facebook page. Temporal Analysis ensures that alarms that are contiguous in time and hence are very likely to be raised due to similar reasons are not forwarded again to the friend(s)-in-loop. It also exploits time information in the recorded audio to reject likely false alarm culprits like music, as we will show later.

The friend(s)-in-the-loop act as a final filter. Humans can be very good at rejecting alarms due to environmental context as they can easily detect the lack of a human scream in audio snippets of reasonable clarity~\cite{ref12}. The friend(s)-in-loop can reduce the overhead of false alarms on law enforcement. They may directly escalate to law enforcement or may confirm to the phone application whether the sample detected as distress is in reality distress or not. Of course, they will introduce delays. Later, we will look at the expected delays between the phone posting an alarm and a friend-in-loop responding to it.

The rest of the paper will describe and evaluate all the blocks (except law enforcement). While we only consider sampling audio information, in case an alarm is confirmed to not be a false alarm, the phone may send any other required information, for example, location, to law enforcement.

\section{Data Collection}
We collected audio that is representative of varied environmental contexts and that captures two expressions of distress, namely screaming and crying. Data collection drives were both controlled and uncontrolled. Uncontrolled data collection involved 16 volunteers recording data using smartphones while going about their daily routines. On the contrary, controlled data collection was done by selected people and was targeted towards collecting specific kinds of audio data. Also, effort was made to get as clean a recording as possible. For example, the collection was done with phone in hand so that recordings were of good fidelity.

All distress audio data was carried out in a controlled manner. We also collected sounds from a set of environmental contexts in a controlled manner. All collected data is recorded at 44.1KHz and is encoded at 16 bits per sample.

\subsection{Controlled Data Collection}
\noindent\textbf{Distress Audio Data:} In this work we consider only screaming and crying. Screams and cry data was collected from the Internet~\cite{ref7}, TV serials, movies, and some data was collected from student actors of age group 18 to 22 years. Silence periods in the beginning and at the end of all recordings were pruned. The pruned audio files were further split into 2 second long audio samples. We have a total of 340 two second long audio samples that represent distress. Out of 340 samples, 315 were from females and the rest few from males.

\noindent\textbf{Normal Human Speech Data:} For normal speech data we collected conversations from TV serials (without background music) and student actors. We have a total of 580 two second samples of normal speech data.

\noindent\textbf{Environmental Context Data:} We sample audio from five different categories. They are described below.
\begin{enumerate}
\item \textbf{Indoors:} This includes the various sounds that are generated when a person is engaged in activities at home and office. Conversations with family members, with people in a meeting at office, dining at home, and any other sounds excluding those generated by equipments and machines found in a typical home or office. We have 8714 two seconds samples in this category. This data was collected by 4 different people using different phones.
\item \textbf{Outdoors:} Outdoors include all sounds when a person is outside. It includes sounds due to vehicular traffic, sounds while commuting in various modes of transport like bus, rail, and car. Other sounds like honking are included. We also were able to collect about 5 minutes of rainfall. In total we have 4513 two second samples under this category. All data was collected using phones.
\item \textbf{Machinery:} This category includes sounds from machines that are commonly used at home and office. Collected sounds include that of Vacuum Cleaners, Hair Dryers, Mixers, Microwave Ovens, Chimney, Flush, Utensils, Shaving Machine, Washing Machine, Rinsing Bowl, fans, air conditioners, and exhausts. The data for this category was collected by a single individual on a mobile and in total we have 3566 two second samples for this category. A small percentage of samples were collected from sources on the Internet.
\item \textbf{TV:} This category includes artificial human sounds and music generated by televisions and radios. We have a total of 3712 two second samples.
\item \textbf{Gathering:} This category includes sounds in situations where many people are together. Large number of people laughing, a crowd cheering, restaurants, malls, markets, seminars, applause by a crowd, and people talking loudly in a metro, are a few examples. This category contains 1641 two-second samples in total. Samples were collected using a phone and from sources on the Internet.
\end{enumerate}

The sounds collected under different categories have small unavoidable overlaps with each other. For example, some of the Home samples had sounds from a TV and music in the background. Similarly, the Indoors category will have human speech data in it.

\subsection{Uncontrolled Data Collection}
Sixteen volunteers (6 females and 10 males) helped us with audio data collection as they went about their daily routines. This data collection was uncontrolled and no instructions about what must be collected were given. The collection was done using smartphones that the volunteers were using as their personal phones. The volunteers were provided with an application that helped them annotate their environmental context. The application nudged the volunteer to annotate as and when a change in sound intensity of greater than 10dB was detected by it. The volunteers had the option of pausing the recording at will. A total of about 250 hours of volunteer data was collected.

\section{Data Processing}
The learning framework cannot process audio samples. Data processing involves converting the audio samples into feature vectors that can be processed by the learning framework. We convert every 2 second audio sample into a 12 element vector that consists of Mel Frequency Cepstral Coefficients (MFCC).

MFCC~\cite{ref27} is widely used in many speech and speaker recognition systems. The mel scale transforms linear frequency into mel-frequency, which represents the frequencies as perceived by the human ear. Small variations in frequencies are not noticed in the mel scale. There are a total of 39 cepstral coefficients. We do not include the 0th cepstral coefficient and coefficients 13 onwards as including them did not give us good classification performance. Our feature vector contains cepstral coefficients 1 to 12.

\section{Learning Framework}
Our supervised learning framework comprises of the algorithms Speech Filter and Context Filter. Both the algorithms use support vector machines to classify any input MFCC vector. Speech Filter uses a SVM model created using the distress-and-speech training set, while Context Filter uses a model created using the distress-and-context training set. The former set consists of the two categories human speech and scream. The latter consists of a total of 7 categories. The creation of the sets is described in detail in Section~7.

Speech Filter is shown in Algorithm~1. It takes as parameters the audio sample that needs to be classified in the form of the corresponding MFCC feature vector, the distress-and-speech model, and a distance measure. The algorithm returns the sample for further escalation if it is categorized as distress. Else, it discards the sample and returns an empty vector. Note that the Speech Filter model contains just the two categories of distress and speech, and thus only one hyperplane~\cite{ref21} that separates speech and distress. As a result the distance estimate $\hat d$ returned by the library function \texttt{svmpredict()} is a scalar. If $\hat d>0$ for the input sample, the function classifies it as distress. Otherwise, it classifies it as normal speech. The absolute value $|\hat d|$ is an indicator of the confidence with which the classification was made.

Speech Filter takes an input parameter, the threshold distance $\DT$, which allows us to choose the values of confidence for which a test sample must be classified as distress. A threshold distance $\DT<0$ will mean that Speech Filter will also accept as distress, test samples that were earlier classified as speech with low confidence. While this will increase the distress detection rate, it will also increase the false alarm rate. Later in Section~8, we show the tradeoff between the detection rate and the false alarm rate (commonly known as the ROC curve~\cite{ref8}) for $-2\leq \DT\leq 4.3$. A suitable choice will be arrived at by running the algorithm on the validation and verifying it on the test data sets.

\par\smallskip\vspace{8.710pt}
\noindent\includegraphics[width=239.103bp]{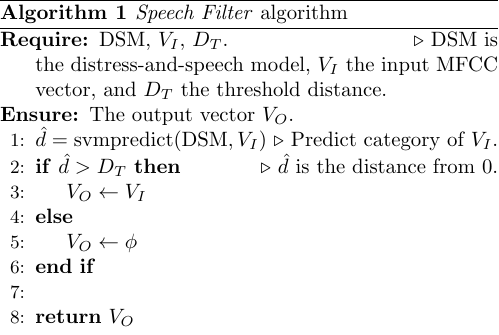}
\par\smallskip\vspace{7.564pt}

Context Filter is described in Algorithm~2. It uses the distress-and-context model that distinguishes between seven categories, one of them being distress. In general, for $K$ categories the model uses $K(K-1)/2$ hyperplanes, one hyperplane separating each pair of categories. Thus, our model has 21 hyperplanes. For every input sample (MFCC vector) the SVM model outputs a predicted category and also a distance measure vector, say $\hat{\vect d}$, of length 21. Assume the categories are labeled from $1,2,\ldots,K$. For ease of exposition, let $K=3$. Then $\hat{\vect d}(1)$ stores the distance measure corresponding to the categories 1 and 2, $\hat{\vect d}(2)$ stores the measure for the categories 1 and 3, and $\hat{\vect d}(3)$ stores the measure for categories 2 and 3.

If a distance measure corresponding to categories $i$ and $j$ is positive, then we say that $i$ wins over $j$, that is the input feature vector is more likely to belong to category $i$ than to category $j$. Else, if the distance measure is negative, $j$ wins over $i$. Having calculated the winning category for each of the $K(K-1)/2$ distance measures, we sort the $K$ category indexes in decreasing order of the number of times they won. Call this sorted list of category indexes as $I$. SVM (\texttt{svmpredict()}) by default returns $I(1)$ as the category of the input sample. That is if $I(1)$ does not store the index \texttt{LABEL\_DISTRESS} of the distress category, SVM will not classify the sample as distress. Context Filter allows for flexibility by using the column relaxation parameter $\CR$, such that if the index of the distress category is amongst the first $\CR$ elements of $I$, it classifies the input as distress. A value of $\CR>1$ therefore allows for a larger distress detection rate at the expense of a larger false alarm rate. If Context Filter classifies an input as distress, it returns the vector, else it returns an empty vector.

Note that, given our set of $K=7$ categories, $\CR$ can take the values $1,2,\ldots,7$. Also, in Speech Filter we will vary the threshold distance $\DT$ over $(-2,4.5)$. When Speech Filter and Context Filter are used together, we end up with a large number of choices that can be made over the set of values in the Cartesian product $\{\CR\}\times\{\DT\}$. We will see in Section~8, that the space allows us to choose a good tradeoff between distress detection rate and false alarm rate. In fact, neither Speech Filter nor Context Filter but the two in series---Context Filter placed after Speech Filter---gives us the best detection rate and FAR tradeoff.

Table~\ref{tab:tradeoff} shows an example of how different choice of $\CR$ and $\DT$ help us choose a better detection rate and FAR tradeoff. For example, $\DT=-1.1$, $\CR=1$ gives a detection rate of 87.38\% and FAR of 1.38\%. This tradeoff can be improved by selecting $\DT=1.9$, $\CR=5$. At this point, the detection rate improves while the FAR is unchanged.

\par\smallskip\vspace{5pt}
\noindent\begin{minipage}{\columnwidth}
\centering
\small
\renewcommand{\arraystretch}{1.087}
\captionsetup{skip=16pt}
\setlength{\tabcolsep}{3.6pt}
\begin{tabular}{c|ccccc}
$\DT/\CR$ & 1 & 2 & 5 & 6 & 7 \\
\hline
$-1.1$ & 1.38 & 2.78 & 9.03 & 14.1 & 57.57 \\
       & 87.38 & 91.18 & 98.88 & 99.82 & 100.00 \\
$0.4$  & 0.28 & 1.01 & 2.75 & 5.08 & 5.72 \\
       & 86.83 & 90.48 & 97.81 & 98.61 & 98.62 \\
$1.9$  & 0.11 & 0.57 & 1.38 & 2.41 & 2.43 \\
       & 85.97 & 88.95 & 95.83 & 96.49 & 96.5 \\
$4.3$  & 0.06 & 0.29 & 0.73 & 1.14 & 1.14 \\
       & 82.61 & 84.67 & 90.70 & 91.09 & 91.1 \\
\end{tabular}
\captionof{table}{An example of how different values of the detection threshold $\DT$ and the column relaxation $\CR$ allow us to choose different detection rate and FAR tradeoff(s). The first row for every value of $\DT$ lists the FAR(s) and the second row lists the detection rates. The values are a subset of the tradeoff(s) in Figure~\ref{fig:roc}b when using In Series learning framework.}
\label{tab:tradeoff}
\end{minipage}
\par\smallskip\vspace{14.6pt}

\begin{figure}[t]
\vspace{0.399pt}
\noindent\includegraphics[width=239.103bp]{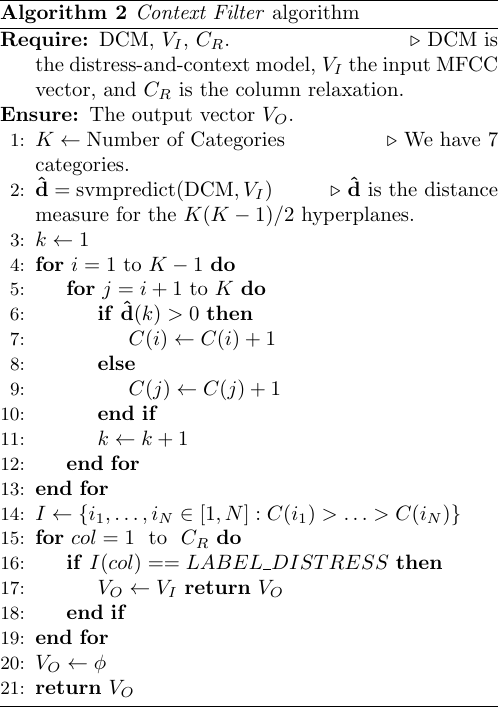}
\vspace{-3.36pt}
\end{figure}

\section{Temporal Analysis}
The temporal analysis (TA) block of \AAsys\ maintains\linebreak a state variable which is set when the learning framework classifies an audio sample as distress. The state variable is unset if no new audio samples are \mbox{classified} as distress for a predefined \texttt{TIMEOUT} period. TA\linebreak has a very simple role. It does not forward any au-\linebreak dio samples classified as scream by the learning frame-\linebreak work to friends-in-the-loop if the state variable is set.\linebreak This ensures, for a reasonable setting of \texttt{TIMEOUT},\linebreak that friends-in-the-loop are not flooded by redundant\linebreak alarms.

\section{Methodology}
The data collected using controlled collection (see Section~3) is used to create two training sets, namely, the distress-and-speech set and the distress-and-context set. The distress-and-speech set includes 50\% of all distress audio samples and also 50\% of all normal human speech samples. In addition, the distress-and-context set includes 50\% of the samples that were collected under the five categories representative of environmental context that we described in Section~3.

The samples that are not a part of the training sets are used to create the validation sets. We create four validation sets. The validation sets are used to find desirable points of operation (the $(\DT,\CR)$ pairs) of our learning framework.

Audio samples that are not part of any training set and belong to human speech and the five environmental context categories are added to each of the four validation sets. Each sample is added as is (its Clean recording), and also at SNR(s) of 40dB, 20dB, and 10dB. A given SNR is achieved for a sample by adding Additive-White-Gaussian-Noise (AWGN) to it in a proportion so as to achieve the said SNR.

The four validation sets differ from one another in the SNR of the distress audio samples added to them. One of the four validation sets includes clean distress samples (no noise added), and the others include distress samples at 40dB, 20dB, and 10dB respectively. Note that unlike the other categories that appear in every validation set at a mix of SNR(s), all distress audio samples in a selected validation set are either clean or have an SNR of 40 or 20 or 10dB.

Also, most importantly, the distress samples are added to samples from the environmental context and human speech and not to AWGN noise to achieve the said SNR. This addition is done in an exhaustive manner. That is, if there are $n$ distress samples and a total of $m$ samples of context and human speech, then we get a total of $nm$ samples of distress each of which is added to one of the $m$ other sounds at the selected SNR.

The different SNR(s) of distress across different validation sets simulate the possibly different relative energies of vocal expression of distress by an individual and that of the surrounding context as sensed by the phone. Note that it is almost impossible to get real data for all possibilities. So we use multiple validation sets to simulate varied environmental contexts, very quiet (high distress SNR) to very harsh (low distress SNR).

\section{Results}
First, we quantify the performance of Speech Filter and Context Filter over our validation sets using different SNR(s) as explained in Section~7. The performance evaluation motivates our learning framework, which has Speech Filter and Context Filter in series. Next, having chosen our learning framework, and a set of desirable column relaxation and distance threshold, $(\CR,\DT)$, pairs, we evaluate the performance of the framework on test data collected by volunteers. This is followed by quantifying the impact of temporal analysis on the false alarm rate obtained from volunteer data.

Note that the learning framework categorizes each input feature vector, which is generated from a 2 sec audio sample. False alarm rate (FAR) is the percentage of such feature vectors, amongst the total non-scream feature vectors, that are wrongly classified as distress. As we go from the learning framework to friends-in-the-loop, we will, instead of stating the FAR, state the average number of posts per hour received by friends-in-the-loop that are false alarms.

\subsection{Validation set evaluation and selection of the learning framework}
Our summary observations are as follows.
\begin{enumerate}
\item Both Speech Filter and Context Filter do well individually when Clean distress audio samples are used.
\item For an SNR of 40dB and less, Context Filter leads to very low distress detection rates.
\item Speech Filter achieves high detection rate. However, the best tradeoff between detection and false alarm rate is achieved when an audio sample is processed by Speech Filter and Context Filter in series. Specifically, every sample is first processed by Speech Filter and in case Speech Filter classifies the input to belong to the distress category, the input is processed by Context Filter. The input is declared to be of the distress category only if Context Filter also classifies it to be distress.
\end{enumerate}

\begin{figure*}[t]
\centering
\vspace{-23.798pt}
\includegraphics[width=520bp]{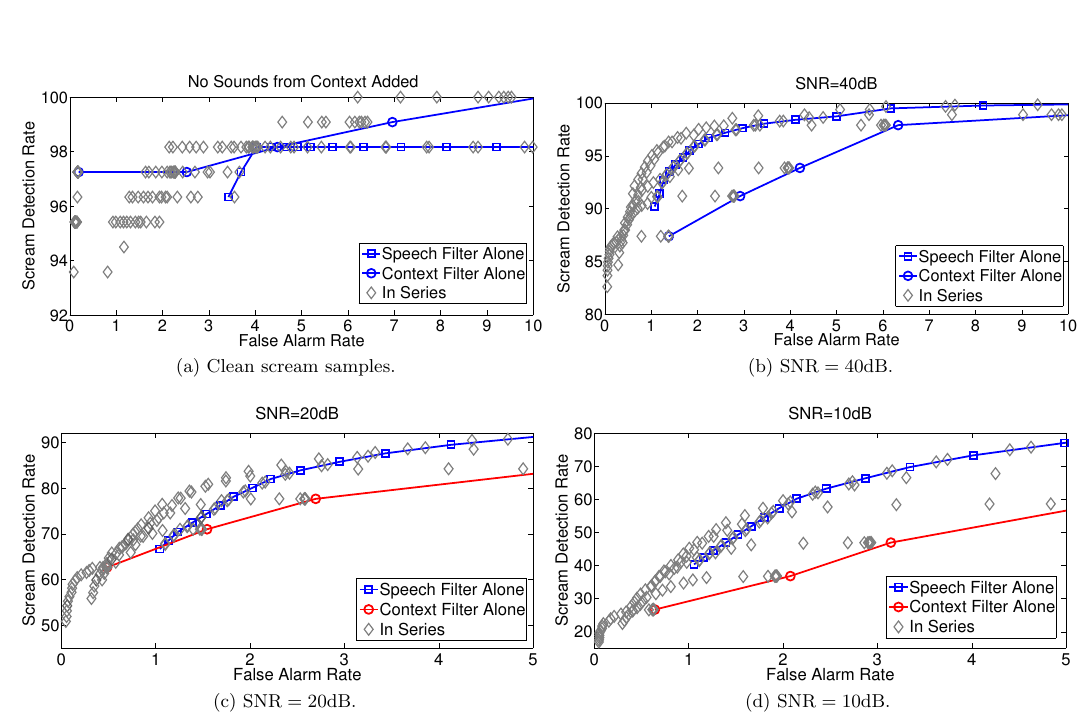}
\caption{The false alarm rate and detection rate tradeoff curves obtained when using each of Speech Filter and Context Filter alone, and In Series. The curves were plotted for distress samples and context related samples in the validation set for distress sample SNR(s) of 10, 20, 40dB, and the case (Clean) when no sounds from the environmental context are added to the distress samples.}
\label{fig:roc}
\end{figure*}

The detection rate and FAR tradeoff(s) achieved are plotted in Figure~\ref{fig:roc}. Figures~\ref{fig:roc}a, \ref{fig:roc}b, \ref{fig:roc}c, and \ref{fig:roc}d, show the tradeoff for distress samples that are Clean (no noise added), have 40dB, 20dB, and 10dB SNR, respectively. For each SNR we show the tradeoff achieved when only Speech Filter is used, only Context Filter is used, and Context Filter follows (is used in series with) Speech Filter. When only Speech Filter is used the different detection rate and FAR tradeoff(s) are obtained by varying the distance threshold $\DT$. When only Context Filter is used, the different tradeoff(s) are achieved by varying the column relaxation $\CR$.

As is seen in Figure~\ref{fig:roc}a, there is not much to choose between the three possibilities when the distress samples are clean. For all other SNR(s) Speech Filter alone achieves much higher detection rates than Context Filter alone. However, note that the best tradeoff between detection rate and FAR is obtained when using the two In Series. For example, consider SNR=40dB (Figure~\ref{fig:roc}b). At about a false alarm rate of 1\%, using In Series can give a detection rate of 5\% more than using Speech Filter alone. Larger gains in detection rate are seen for 20dB SNR (Figure~\ref{fig:roc}c). At SNR=10dB, detection rate is very low. However, for all the SNR(s), the choice of In Series leads to detection rate and FAR tradeoff points that are to the left and top of the points obtained when using Speech Filter and Context Filter alone. Thus, In Series gives a better tradeoff.

\subsection{Choosing a set of desirable points of operation from $\DT\times\CR$}
Each of the tradeoff(s) corresponding to In Series in Figure~\ref{fig:roc} is obtained by a specific setting of $\DT$ and $\CR$. We now choose pairs of values such that a detection rate of about 80\% is achieved at a FAR of about 1\%. For 10dB, a detection rate of 80\% cannot be achieved at a FAR of 1\%. We select two points $P_5$ and $P_6$, where $P_5$ achieves a detection rate of about 80\% at a high FAR of about 6\% and $P_6$ achieves a detection rate of 50\% at FAR of about 1\%. The selected pairs are shown in Table~\ref{tab:points}. The selections for the Clean distress samples are able to achieve FAR of about 1\% even at very high detection rates.

\vspace{18pt}
\subsection{Evaluation over volunteer data}
Our evaluation shows that the false alarm rate is brought down to an average of about one message every three hours using the learning framework and temporal analysis. Assuming that an average individual spends about 12 hours every day outside home, 3 to 4 messages or posts to friends-in-the-loop are the overhead that \AAsys\ creates in lieu for always staying alert to distress.

\begin{table}[t]
\vspace{12pt}
\centering
\small
\setlength{\tabcolsep}{3.0pt}
\renewcommand{\arraystretch}{1.15}
\begin{tabular}{|l|cccccc|}
\hline
& Clean & Clean & 40dB & 20dB & 10dB & 10dB \\
Name & $P_1$ & $P_2$ & $P_3$ & $P_4$ & $P_5$ & $P_6$ \\
\hline
$\DT$ & 4.0 & 3.7 & 2.2 & 1.9 & 0.4 & 2.2 \\
$\CR$ & 3.0 & 2.0 & 4 & 5.0 & 7.0 & 5.0 \\
FAR & 1.29 & 0.99 & 1.02 & 1.35 & 5.71 & 1.29 \\
DR & 96.33 & 95.41 & 95.03 & 79.30 & 79.62 & 50.81 \\
\hline
\end{tabular}
\caption{Selected points of operation $P_1,\ldots,P_6$. For each point we list the value of the distance threshold $\DT$, the column relaxation $\CR$, and the achieved FAR and detection rate. These points are used to evaluate the FAR obtained from volunteer data. We choose two sets of values for Clean and 10dB.}
\label{tab:points}
\end{table}

Note that Table~\ref{tab:points} lists the FAR and detection rate as achieved for the validation data set. We now evaluate performance over the test data set, which consists of data collected by volunteers' phones as they went about\newpage their daily routine. The test data set does not include any real screams.
\vspace{12pt}

\begin{figure}[t]
\centering
\vspace{-8.798pt}
\hspace*{-5.798pt}\includegraphics[width=267bp]{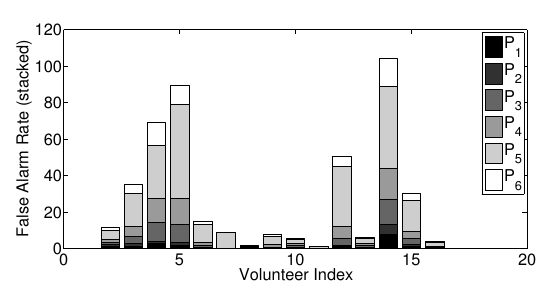}
\caption{FAR experienced by volunteers for the selected points of operation $P_1$ to $P_6$.}
\label{fig:volunteer}
\end{figure}

We evaluate the FAR over volunteer data for the points of operation $P_1$ to $P_6$. The FAR is plotted for each volunteer in Figure~\ref{fig:volunteer}. The FAR for the six points of operation is stacked over one another. For all volunteers, the settings $P_1$ and $P_2$ give very small FAR. $P_3$ works well for some of the volunteers. $P_5$ leads to really large false alarm rates. Note that the variation of FAR with the point of operation is not surprising (also see discussion in Section~11) and was also seen in Figure~\ref{fig:roc}. The higher FAR, for example when using $P_5$, is because the chosen detection rate and FAR tradeoff leads to a very high detection rate, however at a very large FAR. At 40dB SNR, $P_5$ leads to a detection rate of 98.62\% and a FAR of 5.72\% when using In Series. Instead, when $P_3$ is selected we get a detection rate of 95.03\% and FAR of 1.02\%.

For all further evaluation of volunteer data we assume the point of operation to be $P_1$. The median FAR over all volunteers for this choice is 0.45\% and the mean FAR is 0.8978\%. The median and mean drop to 0.58\% and 0.388\% when we exclude volunteer 14. As is seen in Figure~\ref{fig:volunteer}, volunteer 14 generates a large percentage of false alarms, for point of operation $P_1$, in comparison to other users. Our investigation of volunteer 14's recordings revealed music and television to be a big reason for the high FAR. On the flip side, this also leads to the false alarms being contiguous in time. Temporal analysis brings about a big reduction in the number of false alarms that are forwarded to friend(s)-in-the-loop. Row 6 of Table~\ref{tab:temporal} shows the reduction for volunteer 14. A total of 1194 alarms generated by the learning framework are reduced to merely 11.

\subsection{Reduction in FAR on temporal analysis}
Table~\ref{tab:temporal} shows the number of false alarms generated per volunteer assuming contiguous alarm rejection with timeout values of 30, and 60 minutes respectively. If we consider volunteers for whom more than 10 hours of data was recorded, on an average we see a false alarm every 2.5 and 3 hours for timeout settings of 30 min and 60 min respectively.

\begin{table}[t]
\centering
\small
\begin{tabular}{|r|rrr|}
\hline
\multicolumn{4}{|c|}{False alarms sent to friends-in-loop}\\
\hline
Tot. Hours & No Timeout & 30 min & 60 min\\
\hline
57.28 & 304 & 20 & 18\\
39.10 & 55 & 3 & 3\\
19.27 & 466 & 8 & 7\\
18.49 & 25 & 3 & 3\\
18.33 & 7 & 2 & 2\\
13.51 & 1197 & 14 & 11\\
11.09 & 49 & 8 & 6\\
3.15 & 12 & 2 & 2\\
0.97 & 30 & 3 & 2\\
0.69 & 35 & 1 & 1\\
\hline
\end{tabular}
\caption{Each row corresponds to a different volunteer. The first column is the total number of hours of recorded audio from each of the volunteers. The second column lists the number of false alarms sent to friends-in-the-loop assuming that we do not exploit the time contiguous nature of the false alarms. Third and fourth columns show the number of forwarded false alarms for timeout values of 30 and 60 minutes, respectively.}
\label{tab:temporal}
\end{table}

\section{Friends in the Loop}
We chose 3 facebook pages and 2 groups to evaluate the delay between a new post and the first few responses\newpage (comments) to it. This delay is likely to be a good indicator of delays that \AAsys\ will suffer due to the requirement that friends-in-the-loop verify an alarm to be a genuine alarm before it is forwarded to law enforcement. Since the humans that constitute the friends-in-the-loop for any individual is likely to be a small community, the worst case delays may be up to 15 minutes. Note that these delays may reduce in case a directed message is sent to a priori selected group of people, for example, the family of the individual likely to be in distress.

We polled the most recent 300 posts from each page. From each post we obtained the earliest 5 comments. Only posts with at least 1 comment were considered for this experiment. For each post and each of the 5 earliest comments on it, we calculate the difference in the time of posting of the post and the comment. Table~\ref{tab:facebook} summarizes the median of the obtained time differences. Page 3 and page 4 have a very limited number of members and we see median time delays of about 15 minutes between a post and the very first comment on it.

\section{Smartphone Implementation and Energy Considerations}
\begin{figure}[t]
\centering
\hspace*{-1.798pt}\includegraphics[width=248bp]{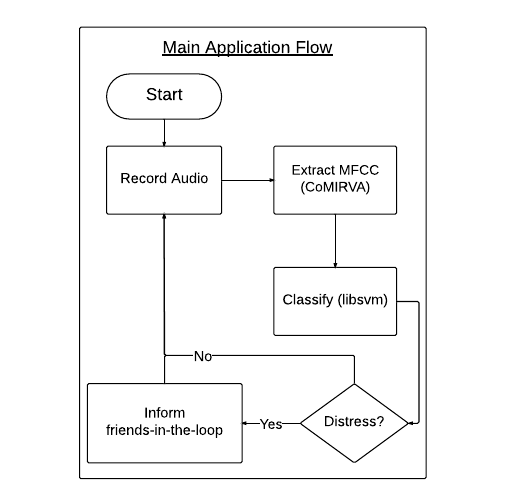}
\caption{The different functional blocks of the \AAsys\ Android application.}
\label{fig:android}
\end{figure}

Figure~\ref{fig:android} summarizes the implementation of the \AAsys\ Android application. We used the CoMIRVA~\cite{ref32} library for MFCC computation. The library libsvm~\cite{ref11} was used for SVM based classification. We now enumerate the challenges faced and the solutions we have implemented.

\begin{table}[t]
\centering
\small
\setlength{\tabcolsep}{2.0pt}
\begin{tabular}{|c|c|c|c|c|c|}
\hline
Comment & Pg. 1 & Pg. 2 & Pg. 3 & Pg. 4 & Pg. 5\\
\hline
1 & 1.233 & 0.892 & 15.52 & 14.48 & 2.2\\
2 & 1.933 & 1.325 & 42.71 & 16.57 & 2.917\\
3 & 2.817 & 1.683 & 80.13 & 22.02 & 3.833\\
4 & 3.65 & 2.025 & 112.6 & 39.43 & 4.542\\
5 & 4.742 & 2.517 & 124.2 & 54.47 & 5.333\\
\hline
\end{tabular}
\caption{For each page, we tabulate the median time between a post and the first to the fifth comment on it. Page 1 has 320000 followers. Page 2 has 3.6 million followers. Page 5 has 68000 followers, page 3 has 622 members, and page 4 has a mere 79 members.}
\label{tab:facebook}
\end{table}

\noindent\textbf{Buffer overflow:} There are two Android libraries, AudioRecord and MediaRecorder, which allow recording audio from the microphone. MediaRecorder only\newpage provides audio in compressed format. As we need to acquire raw PCM data, we used AudioRecord. The AudioRecord API uses the polling approach to acquire PCM frames from the audio buffer, which overflows quickly if not polled. We had to ensure that the thread we created for recording worked in tandem with AudioRecord's thread that polls the buffer so as to ensure that all recorded audio is saved.

\noindent\textbf{Multi-core implementation:} On phones that have multiple cores, we realized that the recording thread was slower than the polling thread since the threads were running on different cores. This led to creation of null frames, which would get saved as a series of 0s and hence corrupt the recording. To make sure these null frames do not get saved, we put each frame through a test to check for null frames.

\begin{figure}[t]
\centering
\vspace{-18.798pt}
\hspace*{-6.812pt}\includegraphics[width=282bp]{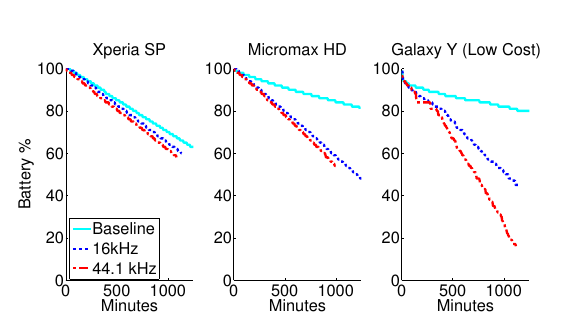}
\caption{Energy consumed for recording at different sampling rates, by three different phones.}
\label{fig:energy}
\end{figure}

\noindent\textbf{User privacy:} We did not want any audio file saved by the application to show up in the media player. Note\newpage that the application may store about 5MB of audio every minute, and the file will need to be saved on the phone's external storage. To hide the audio file we use a combination of three techniques. We do not insert the header into the file, we place a ``nomedia'' marker file in the folder, and make the folder a hidden system folder.

\subsection{Energy consumption}
We measured the energy consumption of recording audio from the microphone and classification using SVM on phone. The experiment was conducted on three Android phones, namely, Samsung Galaxy Y (Android 2.3.6), Sony Xperia SP (Android 4.1.2) and Micromax Canvas HD (Android 4.2.1). Amongst these phones, Galaxy Y is a 2011 model and has low cost hardware, whereas the other two are 2013 models with fairly powerful hardware. The experiments were conducted in two stages, one for recording and the other for classification using libsvm.

Figure~\ref{fig:energy} summarizes the energy consumption during recording. The more powerful phones with newer versions of Android tend to perform better in terms of energy consumption. The sample rate at which the audio was recorded did not create much difference in the energy consumed on Xperia and Micromax. On Galaxy Y it created a huge difference.

The figure does not show the results of the classification experiment. There wasn't a significant difference in terms of energy consumption and classification took merely 3--5\% battery over 10 hours, making classification using libsvm of not much concern with respect to energy consumption. While recording is energy intensive new phones are more optimized for it. In fact, optimizations that improve the audio potential of smartphones are likely to arrive sooner than later.

\section{Limitations and a Few Observations}
\noindent\textbf{On the false alarms:} An average of a post every\newpage three hours can soon become a nuisance, especially if the posts are broadcast to a large community. Also, because almost all of these posts will be false alarms. One may instead choose to send the posts to a smaller group of known people, for example, parents or a spouse. This can lead to privacy concerns, however.

It is worth investigating the ability of the phone to guess the environmental context in which an alarm was raised. Certain contexts may then be treated as safe and alarms raised may be ignored. Of course, the user of \AAsys\ may be nudged and only in case of a previously decided response from the user, the alarm can be discarded by \AAsys.

Throughout this work we have only used inputs from the microphone to detect distress. Inputs from other sensors like accelerometer can help better gauge context and reduce false alarms. GPS or WiFi based location information together with audio fingerprints can help reject repeat false alarms that are tied to a particular location. An example location is a children's park where screaming children lead \AAsys\ to raise false alarms with certainty.

There are other scenarios too. They are not very likely to occur, however, at least not on a daily basis. The framework cannot distinguish between a happy scream (no cause for alarm) and a sad scream, at least not without involving friends-in-the-loop. It cannot currently distinguish between a scream on TV and a real human scream.

\noindent\textbf{Why choose facebook:} We chose facebook to evaluate the delays that friends-in-the-loop may lead to. Chat messengers like WhatsApp will likely lead to smaller delays. However, unlike facebook, we do not have access to API(s) needed to get the required information. Also, our calculations of delays assume that people will over time respond to posts made by \AAsys.

\section{Related Work}
In this section, we summarize related works on distress detection using audio on mobile phones and wearable devices. We will also discuss a few alert dissemination techniques.

\subsection{Distress detection}
Distress detection from atypical events, using audio and video information, is found in~\cite{ref25,ref36,ref37}. Works that only use audio to identify normal and abnormal events include~\cite{ref13,ref18}. For detection of distress, scream is chosen expression of distress~\cite{ref18,ref31,ref35}.

Reference~\cite{ref18} uses the categories of scream, cry, laugh, applause, clap, knock, and glass break to train SVM and GMM with the aim of distinguishing between normal and abnormal events in a home setting. They use a Linux based system (pentium Mobile CPU 2G, 400M front Bus) and a microphone array.

\newpage Reference~\cite{ref35} identifies screams and gunshots using a microphone array and also localizes the abnormal audio event by pointing a camera to the exact location of the event using the Time Difference of Arrival technique. Two parallel GMM(s) were trained to discriminate noise from scream and noise from gunshot. They used a sophisticated feature extraction and selection method to include MFCC along with periodicity, spectral slope and centroid. Reference~\cite{ref31} also detects gunshot, explosion, and scream in a metro and subway environment. The above works provided a good motivation for using scream as a category of interest for our distress detection framework. They discriminated between threatening and non-threatening situations using GMM and HMM. These works show high accuracy for scream detection. However, they have use a dedicated and static hardware for their framework. Also, unlike \AAsys\ they cater to very select environments.

In our work, we carry out a thorough evaluation of scream detection across varied environmental context and different degrees of sounds from the environment. All related works have considered a specific contexts~\cite{ref15,ref22,ref23,ref29,ref31}.

Related works that have used a multi-stage framework include~\cite{ref13,ref15,ref22,ref29,ref30,ref31}. These systems have limited context information, whereas we consider a wide range of possible contexts.

Reference~\cite{ref28} attempts to identify the emotion of the user using GMM and HMM. Emotion Recognition~\cite{ref19,ref20,ref24,ref33} is used to classify six basic type of emotions suggested by Paul Ekman~\cite{ref17}. The emotion recognition work has also been used in fear detection~\cite{ref13}. Their database~\cite{ref14} contains abnormal, non-verbal, and dangerous situations. It was used to make a text-and-speaker-independent fear detection framework. Their accuracy is not high, which is not useful for a system like \AAsys. Emotion recognition is found in other applications as well~\cite{ref9}, but these work within a specific context.

\subsection{Context detection}
Reference~\cite{ref12} emphasizes the need of only audio based systems instead of both video and audio. But the performance shown was limited to non-speech and non-music data, hence a real time deployment of this work is still a question. People have used Audio in Human Activity recognition tasks~\cite{ref22} to recognize illegal activities like trespassing and hunting. The accuracy is very low in presence of noise. Reference~\cite{ref30} uses wavelet transform at its first stage of classification and has considered many context audios. They used a dedicated hardware, however. Audio fingerprints have been used in elderly care to detect falls using mobile and wearable devices~\cite{ref16,ref26}. Reference~\cite{ref26} uses many sensors along with a microphone on a\newpage wearable device to identify depression using silence and speech periods. Reference~\cite{ref34} uses a microphone and other sensors to adapt to user context.

\subsection{Distress Detection using mobile phones}
To our knowledge, there is no work on audio based automated distress detection using mobile phones. Related works on distress detection using mobile phones provide a manual trigger to the user which needs to be pressed when the user is in distress~\cite{ref2,ref3,ref4,ref6}. Reference~\cite{ref2} provides an alert dissemination framework which shows alerts from recent past so that a user is warned in advance about a harmful place.

\section{Conclusions}
We proposed an entirely smartphone audio based $24\times7$ real time distress detection system. We provided an outline of the implementation of the application and a thorough evaluation of its suitability to execute on a smartphone. We proposed a novel framework, that uses two stage learning, temporal information, and ability of humans to reject false alarms based on audio clips to achieve a high distress detection rate and a reasonable false alarm rate. Extensive evaluation on many hours of volunteer data was used to show that the overheads due to false alarm rate are on an average equivalent to a facebook post by the user of \AAsys\ every 3--4 hours.

\balance
\section{References}
\renewcommand{\refname}{}

\end{document}